\documentclass[preprint,showpacs,preprintnumbers,amsmath,amssymb]{revtex4}

\usepackage{graphicx}
\usepackage{dcolumn}
\usepackage{bm}

\begin{document}


\title{Proposal for a geophysical search for dilatonic waves}
\author{Sachie Shiomi}
\email[]{shiomi@faculty.nctu.edu.tw} \affiliation{Space Geodesy
Laboratory, Department of Civil Engineering, National Chiao Tung
University, Hsinchu, Taiwan 300, Republic of China}


\begin{abstract}
We propose a new method of searching for the composition-dependent
dilatonic waves, predicted by unified theories of strings. In this
method, Earth's surface-gravity changes due to translational motions
of its inner core, excited by dilatonic waves, are searched for by
using superconducting gravimeters. This method has its best
sensitivity at the frequency of $\sim$ 7 $\times$ 10$^{-5}$ Hz,
which is lower than the sensitive frequencies of previous proposals
using gravitational-wave detectors: $\sim$ 10 to 1000 Hz. Using
available results of surface-gravity measurements with
superconducting gravimeters and assuming a simple Earth model, we
present preliminary upper limits on the energy density of a
stochastic background of massless dilatons at the low frequency.
Though the results are currently limited by the uncertainty in the
Earth model, this method has a potential of detecting dilatonic
waves in a new window.
\end{abstract}

\pacs{04.80.Cc, 91.10.Pp, 98.70.Vc}

\maketitle

\section{Introduction}

The scalar gravitational fields (or the dilatons) appear in
scalar-tensor theories of gravitation and also in theories toward
the unification of fundamental forces in nature, such as
Kaluza-Klein, supergravity and string theories. Scalar-tensor
theories must agree with general relativity within an accuracy of
$\sim$ 10$^{-4}$ - 10$^{-5}$ in the post-Newtonian limit, because of
the results from the Cassini time-delay experiment
\cite{Bertotti2003} and the search for the Nordtvedt effect (i.e.
violations of the universality of free-fall for massive bodies
\cite{Nordtvedt1968}) \cite{Baessler1999}. However, in strong
gravity systems, in which the post-Newtonian approximation is not
applicable, they could exhibit large observable deviations from
general relativity \cite{Damour1993} and other tests involving a
strong gravity source are necessary. One such tests is to search for
scalar waves from a spherical symmetric gravitational collapse of a
star \cite{Shibata1994, Harada1997}, which are predicted by
scalar-tensor theories but not by general relativity.

In many models of unified theories, the scalar field mediates a new
force (or a fifth force \cite{Fischbach1999}), which leads to a
violation of the universality of free-fall and/or a deviation from
the inverse-square law of gravitation. Such deviations from the
inverse-square law and violations of the universality of free-fall
have been searched for and constrained by various experiments at
different ranges (see e.g. \cite{Fischbach1999,Kapner2007,
Schlamminger2008}). However, string theory, one of the most
promising theories for the unification, predicts the existence of a
relic background of the dilaton, which could be a significant
component of the dark matter in the present universe
\cite{Gasperini2003}; though the fifth-force searches indicate that
the dilaton coupling to matter is very weak, the weakness of the
coupling could be compensated by the high intensity of the
background. Therefore, the detection of the scalar waves or a direct
experimental constraint on the amplitude of the scalar waves would
play an important role for the development of the unified theories
and verification of theory of gravitation.

Motivated by these predictions, strategies of detecting the scalar
waves have been studied (see e.g. \cite{Shibata1994, Bianchi1998,
MaggioreNicolis2000, Gasperini2000, Nakao2001, Gasperini2001,
Babusci2001, Coccia2002, Bonasia2005,Capozziello2006}). These
analyses show that some of the predicted scalar waves could be
detected by using currently operating or advanced planned
ground-based gravitational wave detectors. However, no experimental
results have been presented yet. Here, we propose a new approach of
searching for scalar waves, using superconducting gravimeters, and
we tentatively estimate upper limits on the energy density of a
stochastic background of dilatonic waves. The concept of this method
(described in detail later) is similar to the one of the geophysical
test of the universality of free-fall \cite{Shiomi2006} in the way
that both methods use the Earth as the test body and superconducting
gravimeters as the displacement sensor.

Dilatonic waves can interact with detectors in two ways
\cite{Gasperini1999}: (i) directly, through the effective dilatonic
charges of the detectors, which depend on the internal composition
of the detectors, and (ii) indirectly, through the geodesic coupling
of the detectors to the scalar component of the metric fluctuations.
In the former case, the response of the detectors to the dilatonic
waves is nongeodesic \cite{Gasperini1999}. The proposed method is to
seek for an effect of the direct coupling.

The sensitive frequency range of ground-based gravitational wave
detectors is about 10 to 1000 Hz, while it is about 7 $\times$
10$^{-5}$ Hz for the proposed method; a new window can be explored
with this method. According to a cosmological scenario motivated by
string theory, a relic background of light dilatons, with mass as
small as $\sim$ 10$^{-19}$ eV or less (depending on the level of the
flatness of the background spectrum), could be a significant
component of the dark matter (see Section 6 of \cite{Gasperini2003}
and references therein). The sensitive frequency range of the
proposed method corresponds to the mass range of $\sim$ 3 $\times$
10$^{-19}$ eV; the possible cosmological scenario could be tested
with this method. However, in this paper, we focus on the massless
dilaton for simplicity.

\section{The concept}

In this method, we consider the Earth's inner core as the receiver
of dilatonic waves. The inner core, enclosed in its liquid outer
core, is weakly coupled to the rest part of the Earth mainly by
gravitational forces. When a gravitational wave impinges on the Sun
and the Earth, from any direction that is not aligned with the
Sun-Earth line, their proper separation oscillates. Gravitational
waves couple to matter in the universal way; to first order, the
propagation of gravitational waves would not cause relative motions
between the inner core and the rest part of the Earth. However, when
dilatonic waves pass, because of the difference in dilatonic charge
(namely, the difference in their chemical composition), there would
be relative motions between the inner core and the rest part of the
Earth (see next section for detail). Such relative motions would
result in changes of the gravitational field at the surface of the
Earth. The surface-gravity changes can be searched for by
superconducting gravimeters, which are the most sensitive
instruments for measurements of gravity changes at low frequencies.

The dilatonic charges of the inner core and the rest part of the
Earth are different because of the difference in their chemical
composition (to be discussed in detail later); the inner core mainly
consists of iron and nickel, while the rest part of the Earth is
mainly made of lighter elements such as silicon oxides.

The response of the inner core to the dilatonic waves could be
greatly enhanced when the waves have the same frequencies as the
natural oscillation frequencies of the inner core.

\section{Response of the inner core to dilatonic
waves}\label{st:response}

\begin{figure}
\includegraphics[width=\linewidth]{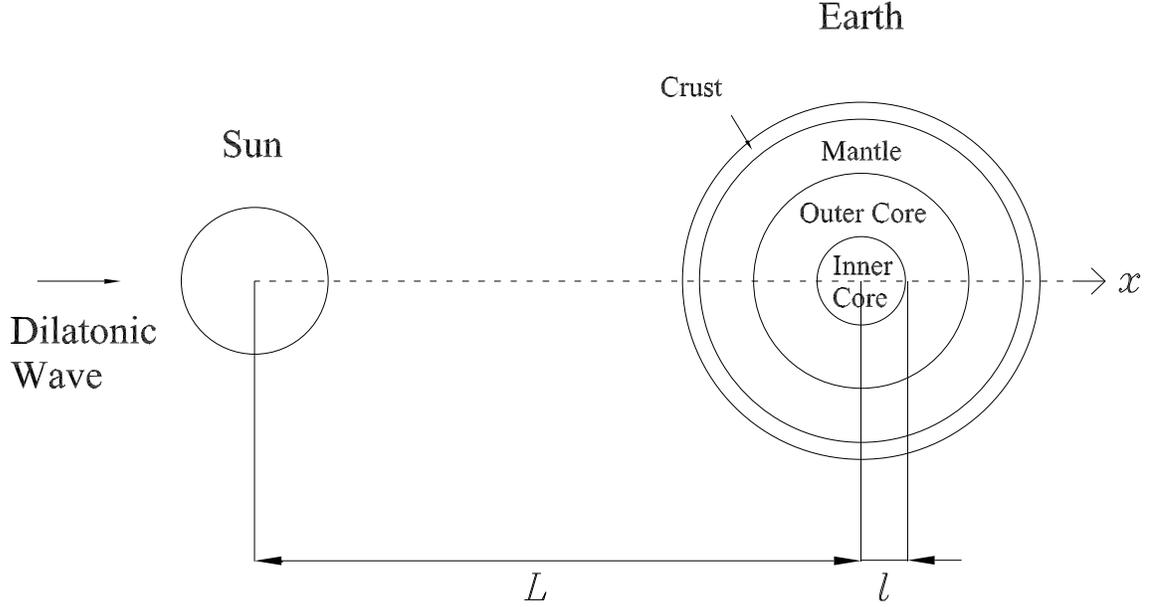}\caption{A schematic cross section
of the assumed configuration (not drawn to scale). As a dilatonic
wave passes along the Sun and the Earth (separated by $L \approx 1.5
\times 10^{11}$ m), the inner core oscillates along the Sun-Earth
line (amplitude $l$). The surface-gravity changes due to the inner
core oscillations are searched for by superconducting gravimeters.
We do not consider the influence of the Earth's rotation on the
inner core for simplicity (see text).} \label{fig:Concept}
\end{figure}

For simplicity, we assume a simple Earth model and configuration as
shown in Fig. \ref{fig:Concept}. A dilatonic wave propagates along
the Sun-Earth line, which connects the centers of mass of the Sun
and the Earth. Their rest separation is $L \approx 1.5 \times
10^{11}$ m.

The Earth's interior can be classified into four parts: the solid
inner core, the liquid outer core, the mantle, and the crust. We
assume that the solid inner core is a homogeneous sphere (with
density $\rho_{ic}$ $\simeq$ 1.29 $\times$ 10$^{4}$ kg m$^{-3}$ and
radius $r_{ic}$ $\simeq$ 1.22 $\times$ 10$^{6}$ m
\cite{Dziewonski1981}) and it is enclosed in the spherical liquid
outer core (with the density of the fluid at the inner core boundary
 $\rho_{oc}$ $\simeq$ 1.22 $\times$ 10$^{4}$ kg m$^{-3}$ \cite{Dziewonski1981}).
We do not consider any deformations (such as tidal and rotational
deformations). We assume that the mantle and the crust are spherical
shells with uniform densities, and their centers of figures are
coincident; their gravitational influence on the inner core is
negligible due to Newton's shell theorem. We do not consider any
electromagnetic effects. Also, we assume that the friction (the
friction coefficient $\gamma$) between the inner core and the outer
core is proportional to the velocity of the inner core.

By applying the generalized equation of geodesic deviation (Eq. (25)
of \cite{Gasperini1999}), which includes the dilaton corrections to
the standard equation of geodesic deviation \cite{Weber1961}, the
translational motion of the inner core, relative to the rest part of
the Earth, can approximately be given by:
\begin{equation}
\ddot{l^i} \approx -\gamma \dot{l^i} - \omega_0^2 l^i - \Delta q
(L^k \partial_k + 1)\partial^i \phi, \label{eq:EofM}
\end{equation}
where $i$ and $k$ indicate the space-time components of the
parameters, $\Delta q$ is the difference in dilatonic charge between
the inner core and the rest part of the Earth (for the definition,
see Eq.(5) below.), $L$ is the rest separation between the Sun and
the Earth (see Fig. \ref{fig:Concept}), and $\phi$ is the dilaton
field. The gravitational stiffness \cite{footnote} and the friction
coefficient are, respectively, given by
\begin{eqnarray} \omega_0^2 &\approx& \frac{4}{3} \pi G
\frac{\rho_{ic} - \rho_{oc}}{\rho_{ic}} \rho_{oc} \approx 1.8 \times
10^{-7} \rm{s^{-2}} \approx \{2 \pi (4.1 \hspace{5
pt}\rm{h})^{-1}\}^2, \label{eq:omega_0^2}\\
\gamma &\equiv& \frac{6
\pi \eta r_{ic}}{m_{ic}} \approx 2.3 \times 10^{-16} \eta \hspace{5
pt} \rm{s^{-1}},
\end{eqnarray}
where $G = 6.67 \times 10^{-11}$ N m$^2$ kg$^{-2}$ is the
gravitational constant, $m_{ic}$ $\approx$ 9.8 $\times$ 10$^{22}$ kg
is the mass of the inner core, and $\eta$ is the effective viscosity
of the outer core. The stiffness due to the Sun's tidal force and
the Moon's tidal force, which act to enlarge any displacement of the
inner core from the center of the Earth, is negligible.

We ignore the influence of the Earth's rotation on the inner core.
The drag associated with the Earth's rotation takes its maximum
value when it is balanced with the Coriolis force ($\sim 2 m_{ic}
\omega_R \omega_0 l_0$, where $\omega_R \approx 2\pi(24
\hspace{5pt}\rm{h})^{-1}$ is the angular frequency of the Earth's
rotation and $l_0$ is a nominal magnitude of the inner-core
displacement), which is about 30 \% of the gravitational restoring
force ($m_{ic}\omega_0^2 l_0$); for the liquid outer core, the drag
would be less than the maximum value. The central force should be
less than 2 \% of the gravitational restoring force.

The last term of Eq. (\ref{eq:EofM}) is due to the direct coupling
to the gradients of the dilaton field $\phi$. $\Delta q$ is the
difference in dilatonic charge between the inner core and the rest
part of the Earth. For ordinary matter, the dilatonic charge can be
obtained by summing over all the components $n$ of the matter
\cite{Gasperini1999}: $q = \frac{\sum_n m_n q_n}{\sum_n m_n}$; $q$
is defined as the relative strength of scalar to gravitational
forces. For a body (mass $M$) composed by $B$ barions with mass
$m_b$ and (dimensionless) fundamental charge $q_b$, and $Z$ ($\simeq
B$) electrons with mass $m_e$ and (dimensionless) fundamental charge
$q_e$, we obtain (Eq. (22) of \cite{Gasperini1999})
\begin{equation}
q\simeq \frac{B m_b q_b}{M} \equiv \frac{B}{\mu}q_b,
\end{equation}
where $\mu = \frac{M}{m_b}$ is the mass of the body in units of
baryonic masses. Therefore, $\Delta q$ can be given by
\begin{equation}
\Delta q \simeq \left\{\left(\frac{B}{\mu}\right)_{ic} -
\left(\frac{B}{\mu}\right)_{rpe}\right\}q_b \equiv \Delta
\left(\frac{B}{\mu}\right)q_b,
\end{equation}
where \textit{ic} and \textit{rpe} denote the inner core and the
rest part of the Earth, respectively. The magnitude of $\Delta
\left(\frac{B}{\mu}\right)$ is about 10$^{-3}$ or less (see e.g.,
Table 2.1 of \cite{Fischbach1999}). The upper limit on $q_b^2$ is
determined to $\sim$ 1.5 $\times$ 10$^{-10}$ by tests of the
equivalence principle \cite{Schlamminger2008}. Therefore, we obtain
$\Delta q \lesssim 1 \times 10^{-8}$.

When a dilatonic wave propagates along the Sun-Earth line, the
spectrum of the displacement along the line is given by
\begin{equation}
S_l (f_0) = \frac{(\Delta q)^2 (L^2 \kappa^2 +1)\kappa^2
c^4}{4\omega_0^2 \gamma^2} S_h(f_0) \label{eq:S_l}
\end{equation}
at the resonant frequency: $f_0 \equiv \omega_0/2 \pi \approx 6.8
\times 10^{-5} \textrm{ Hz}$, where $S_h(f)$ is the power spectrum
of the strain due to massless dilatonic waves: $h \equiv 2\phi/c^2$;
$\kappa$ is the wave number: $\kappa = \omega_0/c$; and $c$ is the
speed of light.

The intensity of a stochastic background of massless dilaton can be
characterized by the dimensionless energy density
\cite{Gasperini2003}: $\Omega_{DW}(f) \equiv (1/\rho_c)(d\rho_{DW}/d
\log f$), where $\rho_c = 3 H_0^2 c^2(8 \pi G)^{-1}$ is the present
value of the critical energy density for closing the Universe; $H_0$
is the Hubble expansion rate: $H_0 = 3.2 \times 10^{-18} h_{100}$
s$^{-1}$; $h_{100}$, ranging from $\sim$ 0.5 to 1, is a
dimensionless factor to account for different values of $H_0$.

The power spectrum of the strain is related to the dimensionless
energy density by the following expression \cite{Bonasia2005}:
\begin{equation}
\Omega_{DW}(f) = \frac{\pi^2 f^3}{3 H_0^2} S_h(f)\label{eq:Omega_DW}
\end{equation}
for massless dilaton. Here the stochastic dilatonic background is
assumed to be isotropic, unpolarized, and stationary
\cite{Bonasia2005}.

\section{Estimation of the upper limits}\label{st:constrains}

Coriolis acceleration splits the inner core oscillation into a
triplet of periods (the Slichter triplet \cite{Slichter1961}). To
determine physical properties of the core, the three translational
mode signals (prograde equatorial, axial, retrograde equatorial)
have been searched for in geophysics (e.g. \cite{Smylie1992,
Hinderer1995, Courtier2000, Rosat2003, Rosat2004, Sun2004,
Rosat2006, Guo2007}), but the detection has not been confirmed yet.
We use a recent analysis by Rosat {\it et al.} \cite{Rosat2004} to
estimate the upper limit on the magnitude of the inner-core
displacement.

Rosat {\it et al.} examined gravity data obtained over the year 2001
from global observatories (in Canada, Australia, Japan, France, and
South Africa) by applying a multistation stacking method
\cite{Rosat2004}. They found no significant peaks that could
originate from the translational motions of the inner core. The
average noise level and the standard intervals were (4.2 $\pm$ 1.6)
$\times$ 10$^{-12}$ m s$^{-2}$ Hz$^{-1/2}$ at $\sim$ 7 $\times$
10$^{-5}$ Hz in their analysis (Fig. 12(b) of \cite{Rosat2004}).
This indicates that the upper limit on the amplitude of each
translational mode is $\delta g =$ 5.8 $\times$ 10$^{-12}$ m
s$^{-2}$ Hz$^{-1/2}$ at the frequency. This upper limit
approximately corresponds to a displacement of the inner core:
\begin{eqnarray}
x \approx \frac{r_e^3}{2Gm_{ic}} \delta g \approx 0.11 \textrm{ mm
Hz}^{-1/2},
\end{eqnarray}
where $r_e = 6371$ km \cite{Dziewonski1981} is the mean radius of
the Earth.

We assume that $x$ represents the upper limit on the typical
magnitude of the displacement. With the value of $x$ and Eq.
(\ref{eq:S_l}), we obtain the upper limit on the strain spectrum:
\begin{equation}
\sqrt{S_h(f_0)} \leq \frac{2 \gamma}{\Delta q \cdot c \sqrt{L^2
\omega_0^2 /c^2+1}} \sqrt{S_l(f_0)} \approx 1.5 \times 10^{-20} \eta
\textrm{ Hz}^{-1/2}.\label{eq:sqrt_Sh_limit}
\end{equation}

\begin{figure}
\includegraphics[width=\linewidth]{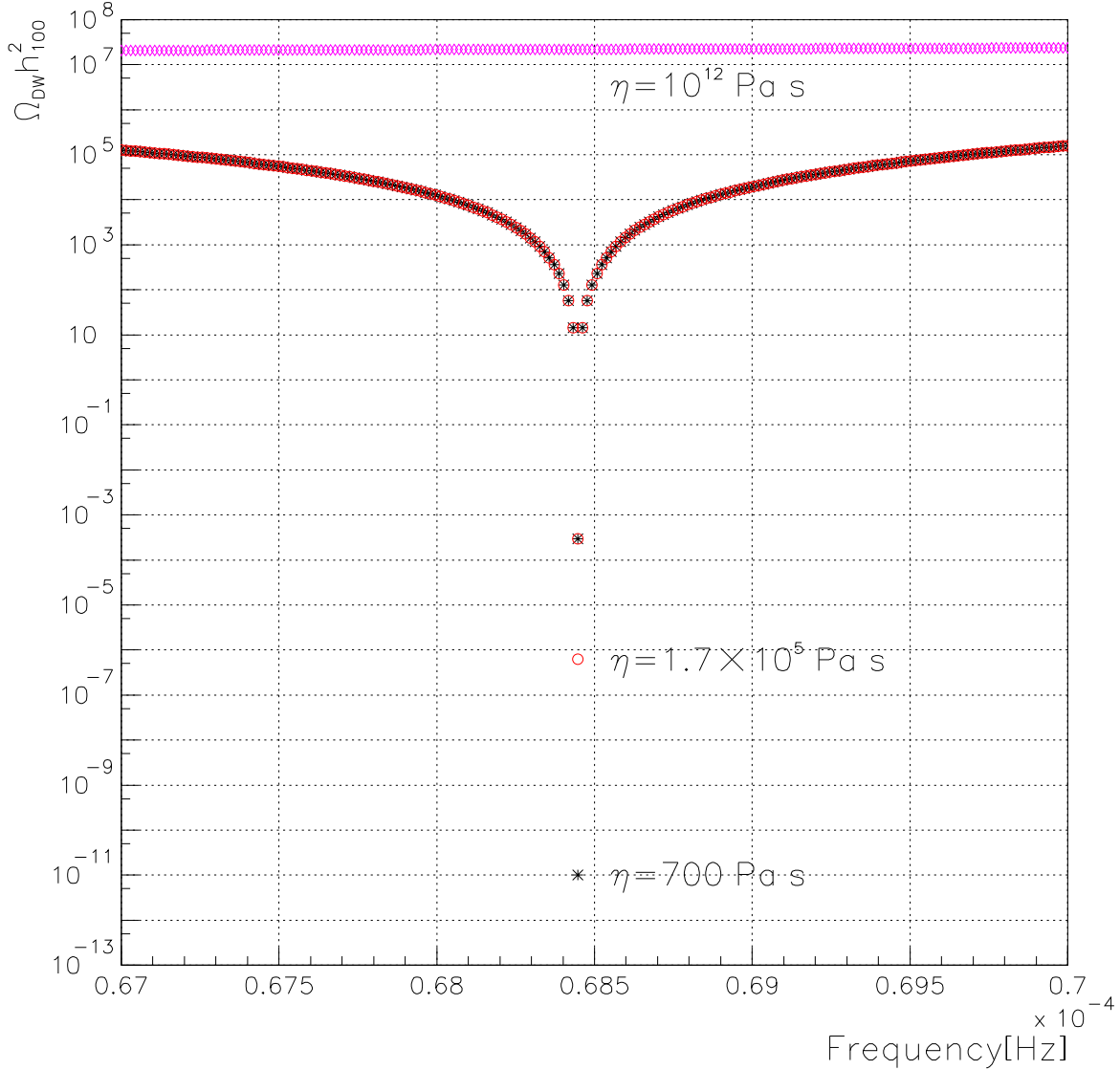}\caption{Expected upper limits on $\Omega_{DW}h_{100}^2$ for different values of $\eta$.
Asterisks, circles and diamonds indicate expected upper limits for
$\eta$ = 700 Pa\,s (when the Reynolds number is unity),
1.7$\times$10$^5$ Pa\,s (the upper bound estimated from nutation
data \cite{Mathews2005}), and 10$^{12}$ Pa\,s (the largest estimate
given in \cite{Secco1995}), respectively (see text).}
\label{fig:Omegah2}
\end{figure}

From Eq. (\ref{eq:Omega_DW}) and the upper limit on the strain
spectrum, we obtain
\begin{equation}
\Omega_{DW}(f_0)h_{100}^2 < 2.2 \times 10^{-17} \eta^2.
\label{eq:Omega_DW_limit}
\end{equation}
The effective viscosity $\eta$ is not well determined and estimates
from various methods vary from $\sim$ 10$^{-3}$ Pa s to $\sim$
10$^{12}$ Pa s \cite{Secco1995}. The Reynolds number ($\equiv
\rho_{oc} v r_{ic}/ \eta$, where $v$ is the velocity of the inner
core) is less than unity when $\eta$ is larger than $\sim$ 7
$\times$ 10$^2$ Pa s and the amplitude of oscillations $x$ ($\equiv$
$v/\omega_0$) is 0.11 mm. We obtain $\Omega_{DW}(f_0)h_{100}^2 <
1\times 10^{-11}$ when the Reynolds number is unity and
$\Omega_{DW}(f_0)h_{100}^2 < 2 \times 10^{7}$ in the worst case when
the effective viscosity is as large as $10^{12}$ Pa s. When we use a
recent estimate of an upper bound from nutation data
\cite{Mathews2005}, $\eta \leq$ 1.7 $\times$ 10$^5$ Pa s, we obtain
$\Omega_{DW}(f_0)h_{100}^2 < 6\times 10^{-7}$. The expected upper
limits for those different values of $\eta$ are shown in Fig.
\ref{fig:Omegah2}.

\section{Discussion}
By assuming the simple Earth model and using available geophysical
results, we have estimated the upper limits on the dimensionless
energy density of the stochastic background of massless dilaton. As
one can see in the relation (\ref{eq:Omega_DW_limit}), the upper
limits largely depend on the magnitude of the effective viscosity of
the outer core, which is poorly constrained by experiments.
Identification of the magnitude of the friction coefficient is
crucial to this method.

There are a number of active researches going on to determine the
effective viscosity and other physical parameters of the Earth's
interior. Some of them are laboratory measurements of the viscosity
at high pressures and temperatures \cite{Mineev2004}, studies for
neutrino oscillation tomography of the Earth's interior
\cite{Winter2005}, and coincidence measurements with a laser
strainmeter system and a superconducting gravimeter at the Kamioka
Observatory in Japan \cite{Takemoto2004}. With these researches
employing new technologies, one can expect that our knowledge on the
effective viscosity and dynamics of Earth's interior will be
improved significantly in the near future.

The dimensionless energy density of a massless background is
constrained to $\Omega h_{100}^2 \lesssim 10^{-5}$, in the frequency
range of our interest, by the nucleosynthesis and resent
measurements of the cosmic microwave background \cite{Cyburt2005}.
From the relation (\ref{eq:Omega_DW_limit}), one can see that, if
the effective viscosity is smaller than about 7 $\times$ 10$^{5}$ Pa
s, the current sensitivity of this method is sufficient to reach the
limit imposed by the astrophysical observations. As for the relic
background of light dilatons, mentioned in the introduction, the
intensity of the background is not constrained by the astrophysical
observations and could compensate the weakness of the dilaton
coupling \cite{Gasperini2003}. The possibility of testing the
cosmological scenario has to be discussed in the future.

We have assumed that the dilatonic waves are isotropic. However, if
they are not isotropic, the magnitude of the inner-core displacement
would depend on the location of the observatories. Such location
dependence could be investigated through the Global Geodynamics
Project network (GGP \cite{GGP}) of superconducting gravimeters.

Though this method is currently limited by the uncertainty in the
Earth model, it is in principle sensitive to any
composition-dependent waves at low frequencies. Other possible
sources of such waves and their influence on the inner core have to
be investigated.

Geophysical sources of the excitation of translational motions of
the inner core are not well known. Possible sources of the
excitation are, for example, large earthquakes \cite{Crossley1992},
some dynamic flows \cite{Greff-Lefftz2007}, and magnetohydrodynamic
processes in the core \cite{Kuang2006}. The effects of earthquakes
are thought to be very small \cite{Crossley1992}. If the
translational motions are experimentally confirmed, all possible
sources of the excitation have to be considered carefully before
estimating the amplitude of the dilatonic waves.

The sensitivity could be improved, for instance, by extracting
correlated signals from the GGP network and applying an advanced
data analysis method \cite{Rosat2008}. Also, the measurement
sensitivity could be improved by further studies.

\section{Conclusions}

We have proposed a new method of searching for dilatonic waves. In
this method, Earth's surface-gravity changes due to translational
motions of the inner core, excited by dilatonic waves, are searched
for by using superconducting gravimeters. The main merits of this
method may be as follows:(1) unlike the previous proposals, which
intend to use gravitational wave detectors, this method employs the
geophysical approach with superconducting gravimeters, (2) the
sensitive frequency range is low ($\sim$ 7 $\times$ 10$^{-5}$ Hz),
in comparison with the previous proposals ($\sim$ 10 to 1000 Hz),
(3) it is devoted to searching for the direct coupling of dilatonic
waves to matter, and (4) it has a potential to search for anisotropy
in composition-dependent waves, by observing anisotropy in the
oscillation direction of the inner core through the GGP network. The
major obstacle of this method is its dependence on the Earth model.
This obstacle would diminish in time with the progress on
understanding of the Earth's interior.

Assuming the simple Earth model and configuration, we have obtained
the preliminary upper limits on the stochastic massless dilatonic
background at the frequency of $f_0$ $\approx$ 7 $\times$ 10$^{-5}$
Hz. In the worst case when the effective viscosity ($\eta$) is as
large as 10$^{12}$ Pa s, the upper limit on the dimensionless energy
density ($\Omega_{DW}(f_0)h_{100}^2$) is 2 $\times$ 10$^{7}$. When
the Reynolds number is about unity ($\eta \simeq$ 7 $\times$ 10$^2$
Pa s), the upper limit is 1 $\times$ 10$^{-11}$. With the recent
estimate of the upper bound from nutation data, $\eta \leq$ 1.7
$\times$ 10$^{-5}$, we obtain $\Omega_{DW}(f_0)h_{100}^2 \leq 6
\times$ 10$^{-7}$.

This method is in principle sensitive to any composition-dependent
waves at low frequencies. Other possible sources of such waves have
to be investigated. We have focused on the massless dilaton. The
possibility of searching for the relic background of light dilatons,
predicted as a scenario in string cosmology, has to be discussed in
the future.

The sensitivity of this method could be increased with further
studies, such as a development of coincidence measurements through
the GGP network. With the increased sensitivity and refined Earth
model, we could test predictions of unified theories and
cosmological models in a new window.

\begin{acknowledgments} The author would like to thank H. Hatanaka for helpful discussions,
A. Pulido Pat\'{o}n for his comments on the manuscript, and C. Hwang
and colleagues for their encouragements and support throughout this
work. The author would also like to thank the referees for their
useful comments and suggestions. This work is funded by the Ministry
of the Interior of Taiwan, under a superconducting gravimeter
project.
\end{acknowledgments}

\bibliography{apssamp}

\end{document}